\documentclass[prl,aps,amssymb,shownopacs,twocolumn,10pt]{revtex4-1}
\usepackage{amsmath}
\usepackage{amssymb}
\usepackage{amsthm}
\usepackage{amsfonts}
\usepackage{listings}
\usepackage{enumerate}
\usepackage{latexsym}

\usepackage{psfrag}

\usepackage{bm}
\usepackage[pdftex]{graphicx}
\usepackage{subfigure}

\newcommand{\beq}{\begin{equation}}
\newcommand{\eneq}{\end{equation}}

\newcommand{\bra}[1]{\left\langle#1\right|}
\newcommand{\ket}[1]{\left|#1\right\rangle}

\def\be{\begin{equation}}
\def\ee{\end{equation}}
\def\ba{\begin{eqnarray}}
\def\ea{\end{eqnarray}}

\input{epsf}

\begin{document}

\tolerance 10000

\newcommand{\vk}{{\bf k}}

\title{Matrix Product States for Trial Quantum Hall States}

\author{B. Estienne$^1$, Z. Papi\'c$^2$, N. Regnault$^{2,3}$, and B. A. Bernevig$^2$}
\affiliation{$^1$LPTHE, CNRS, UPMC Univ Paris 06 Bo\^ite 126, 4 Place Jussieu, F-75252 PARIS CEDEX 05}
\affiliation{$^2$Department of Physics, Princeton University, Princeton, NJ 08544}
\affiliation{$^3$Laboratoire Pierre Aigrain, ENS and CNRS, 24 Rue Lhomond, 75005 Paris, France}

\begin{abstract}
We obtain an exact matrix-product-state (MPS) representation of a large series of fractional quantum Hall (FQH) states in various geometries of genus $0$. The states in question include all paired $k=2$ Jack polynomials, such as the Moore-Read and Gaffnian states, as well as the Read-Rezayi $k=3$ state.  We also outline the procedures through which the MPS of other model FQH states can be obtained, provided their wavefunction can be written as  a correlator in a $1+1$ conformal field theory (CFT). The auxiliary Hilbert space of the MPS, which gives the counting of the entanglement spectrum, is then simply the Hilbert space of the underlying CFT. This formalism enlightens the link between entanglement spectrum and edge modes. 
 Properties of model wavefunctions such as the thin-torus root partitions and squeezing are recast in the MPS form, and numerical benchmarks for the accuracy of the new MPS prescription in various geometries are provided.
\end{abstract}

\date{\today}

\maketitle

The understanding and simulation of quantum many-body states in one space dimension has experienced revolutionary progress with the advent of the density matrix renormalization group~\cite{White-PhysRevLett.69.2863}. In modern language, this method can be viewed as a variational optimization over the set of matrix product states (MPS)~\cite{fannes1992finitely, perez-garcia}. Indeed, gapped one-dimensional systems (which generally have low entanglement) can be very efficiently simulated by expressing the weights of  many-body non-interacting states in an interacting wavefunction as products of finite-dimensional matrices $B[i]^{m_i}$ associated with each occupied ($m_i=1$) or unoccupied ($m_i= 0$) site $i$ (or spin): $\ket{\psi}= \sum_{\{m_i\} }P_L B[1]^{m_1} B[2]^{m_2} \ldots  B[N_s]^{m_{N_s}} P_R \ket{m_1, m_2,\ldots m_{N_s}}$. $P_L, P_R$ are projectors into the state at the end of the chain (absent for periodic boundary conditions or the infinite chain). As long as the "bond dimension" $\chi$ of the matrix $B[i]$ is less than $2^i$, this provides a more economical representation of the state. Generic 1-D gapped systems can be approximated by finite $\chi$~\cite{verstraete}. Critical systems however require an MPS with an infinite bond dimension \cite{Cirac-PhysRevB.81.104431,Nielsen-JPhysStat-P11014}.

Due to their perimeter law entanglement, 2-D systems (such as the fractional quantum Hall effect) are harder to simulate by MPS ~\cite{ashvin, balents-2012NatPh.8.902J, cirac, vidal-PhysRevLett.110.067208}. In a recent paper \cite{zaletel-PhysRevB.86.245305}, exploiting the fact that continuum model FQH states can be written as correlators of primary fields in conformal field theories (CFTs), an MPS expression was obtained for continuum Laughlin~\cite{laughlin-PhysRevLett.50.1395} and Moore-Read~\cite{Moore1991362} states on infinite cylinders. The bond dimension $\chi$ grows with the number of particles but scales with the circumference $L$ of the cylinder rather than its area.  Approximate expressions can be obtained by truncating $\chi$ of the exact MPS. A key ingredient of Ref.~\cite{zaletel-PhysRevB.86.245305} is the expansion of operators in a free basis (boson for Laughlin and Majorana plus boson for Moore-Read), which cannot be easily implemented in the more complicated, interacting bases of other FQH states such as the Read-Rezayi series~\cite{read-PhysRevB.59.8084}. The exciting possibility is that, if all model FQH states could be written in MPS form, current numerical barriers could be broken and properties such as correlation functions would be computable for large sizes. This is supported by the continuous MPS proposed in Ref.~\cite{dubail-PhysRevB.86.245310}.

In this paper we provide a generic prescription that enables us to obtain the unnormalized (thin annulus) MPS form of a model FQH state which is the correlator of a primary field in a CFT. We explicitly construct MPS for the $(k,r)=(2,r)$ paired Jack polynomial states~\cite{Bernevig08:Jack,Bernevig08:Jack2} ($r=2,3$ being the Moore-Read and Gaffnian~\cite{simon2006} wavefunctions), as well as for $(k,r)=(3,2)$ corresponding to the Read-Rezayi $Z_3$ wavefunction~\cite{read-PhysRevB.59.8084}. Several key ingredients and subtleties, such as the presence of non-orthonormal bases, null vectors, and intricate operator commutation relations, are discussed. We then show how to extend the MPS description to different manifolds such as the cylinder, sphere, and plane, and how several known properties of the CFT wavefunctions such as squeezing arise naturally in this description. We then generate several (the $(k,r)= (2,2), (2,3), (2,6), (3,2)$) of these states numerically to verify our MPS, and provide numerical benchmarks to attest the accuracy of the MPS on the cylinder\cite{rezayi-PhysRevB.50.17199} and sphere\cite{haldane-PhysRevLett.51.605}.

A large class of FQH ground-states are described by the  many-point correlation function of an electron operator field $V(z)$ in a chiral $1+1$ CFT~\cite{Moore1991362}: 
\begin{align}
\langle N_e \sqrt{q}|  V(z_{N_e}) \ldots V(z_{1}) | 0\rangle  = \sum_{\lambda} c_{\lambda} m_{\lambda}(z_1,\ldots,z_{N_e}) \label{CFT_correlator}
\end{align}
where $m_{\lambda}$ are monomials (or Slaters for fermions) of angular momentum $\lambda= (\lambda_1,\ldots, \lambda_{N_e})$. $N_e$ is the  number of electrons, and the filling fraction is $\nu =1/q$ (note that $q$ will not always be integer). The state $\langle N_e \sqrt{q}|$ describes the background charge at infinity.  The coefficient $c_{\lambda}$ can be obtained by contour integrals. Upon inserting a complete basis of states in the l.h.s. of \eqref{CFT_correlator} we get
\begin{align}
c_{\lambda} = \sum_{\{\alpha_j\}}   \prod_{j=1}^{N_e} \frac{ 1 }{ 2 \pi i  } \oint \frac{\text{d}z_j}{{z_j^{\lambda_j+1} } }  \langle \alpha_{j} | V(z_j) | \alpha_{j-1} \rangle. 
\end{align}
The $U(1)$ charge of $\ket{\alpha_0}, \ket{\alpha_{N_e}}$  is $0, N_e \sqrt q$, respectively.  This is  an infinite, site (Landau level orbital momentum) dependent MPS $|\Psi \rangle = \sum_{\{ m_i \}} \left( \tilde{B}^{m_1}[1] \ldots \tilde{B}^{m_{N_{\Phi}}}[N_{\Phi}]\right)_{N_e \sqrt{q}, 0} | m_1 \ldots m_{N_{\Phi}} \rangle$, 
with the matrices for an orbital $j$ being, in the limit of an annulus with a very large radius (the so-called "conformal limit"),   $\langle \alpha' | \tilde{B}^0[j] | \alpha \rangle  = \delta_{\alpha',\alpha} $  and 
$   \langle \alpha' | \tilde{B}^1[j] | \alpha \rangle   = \delta_{\Delta_{\alpha'},\Delta_{\alpha}+h+j} \langle \alpha' | V(1) | \alpha \rangle $. $h$ is the conformal dimension of $V(z)$, and higher occupation number (of occupation $m$) matrices are simply $ \tilde{B}^m[j]  =   \left(\tilde{B}^1[j]\right)^m/\sqrt{m!}$.

To obtain a site-independent MPS,  we need to spread the background charge uniformly over the droplet. We make explicit the dependence on the $\text{U}(1)$ charge by writing states $|\alpha \rangle = |Q\rangle \otimes |\tilde{\alpha} \rangle$, where $Q$ is the $U(1)$ charge and $\tilde{\alpha}$ encodes the rest (descendant, neutral sector). As the matrix element $\langle \tilde{\alpha}' | V(1) | \tilde{\alpha} \rangle $ does not depend on charge $Q$, we are free to modify the distribution of the background charge. Spreading uniformly the background charge amounts to an insertion of a $U(1)$ background charge $-1/\sqrt{q}$ between each orbital.
This yields a site independent MPS with $ B^m  = e^{-i/2\sqrt{q}\varphi_0}V_0^m e^{-i/2\sqrt{q}\varphi_0}/\sqrt{m!}$, where $\varphi_0$ is the $U(1)$ boson zero mode. 
\begin{align}
\langle \alpha' |V_0 | \alpha \rangle = \frac{1}{2\pi i } \oint \frac{\text{d}z}{z} \, \langle \alpha' |V(z)   | \alpha \rangle,
\end{align}
where $\alpha$ and $\alpha'$ are the basis of descendants in the CFT (not free fermions as in Ref.~\onlinecite{zaletel-PhysRevB.86.245305}).  Our expression further differs from the one  in Ref.~\onlinecite{zaletel-PhysRevB.86.245305} by time-evolution terms $U(\delta \tau) = \exp ( - \delta \tau L_0)$, which give the cylinder normalization. Conformal invariance yields 
\begin{align}
\langle \alpha' |V(z)   | \alpha \rangle =  z^{\Delta_{\alpha'}- \Delta_{\alpha}-h} \langle \alpha' | V(1) | \alpha \rangle, \label{conformalinvarianceandV}
\end{align}
where $h,\Delta_\alpha$ are the conformal dimensions of the primary field $V(z)$ and of the descendent $\ket{\alpha}$. The matrix elements of $V_0$ are then related to the CFT 3-point function $\langle \alpha' |V_0   | \alpha \rangle =  \delta_{\Delta_{\alpha'}, \Delta_{\alpha}+h} \langle \alpha' | V(1) | \alpha \rangle$. The "electron operator" is a primary field of the tensor product form $V(z) =    \Phi(z)  \, \otimes\, :e^{i\sqrt{q}\varphi (z)}: $, where $\Phi(z)$ lives in the so-called neutral $| a \rangle $ conformal field theory $\text{CFT}_n$  factorized from the $U(1)$ sector $ |b \rangle$. In this basis  $ |\alpha \rangle = | a \rangle \otimes |b \rangle$ the 3-point function factorizes as $\langle a';b'| V(1) | a ;b \rangle =
 \delta_{\Delta_{a'}+\Delta_{b'},\Delta_{a}+\Delta_{b}+h}   \langle a' | \Phi(1) | a \rangle \, \, \langle  b' | :e^{i \beta \varphi(1)}: |b \rangle \label{V_0 factorized} $
where $h = q^2/2 +h_{\Phi}$ is the conformal dimension of $V(z) $. Note the delta function in the total conformal dimension of the field and not separately in the neutral and $U(1)$ parts.  In the following we explain how to obtain the neutral, interacting CFT matrix elements for the case of $(k,r)$ Jacks for $k=2$ and $k=3$.

First, we re-evaluate the $U(1)$ matrix elements in a way easily generalizable to non-free field CFT and in a basis where they are real:
\begin{align}
\varphi(w) = \varphi_0 - i a_0 \log (w) +  i\sum_{n\neq 0} \frac{1}{n}a_nw^{-n},
\end{align}
where $a_n$ are the bosonic modes obeying the Heisenberg algebra $ [a_n,a_m] = n\delta_{n+m,0}$. $a_0$  is the zero mode of the conserved current and measures the $\text{U}(1)$ charge, while $\varphi_0$ is its canonical conjugate ($[\varphi_0,a_0]=i$). Primary fields are the vertex operators $\mathcal{V}_{\beta}(z) = :e^{i\beta \varphi(z)}: $ with conformal dimension  $\beta^2/2$. The corresponding highest weight state $|\beta \rangle = \mathcal{V}_{\beta}(0) |0\rangle$, which is  annihilated by all $a_{n>0}$, has $U(1)$ charge $\beta$. Since we defined the $U(1)$ charge to be $a_0$, the electric charge is $\frac{1}{\sqrt{q}}a_0$. Descendants are obtained  by acting on $|\beta \rangle$ with the lowering operators $a_n^{\dagger}=a_{-n}$, $n>0$. They are labeled by a partition $\mu= \{\mu_j\}$, with $\mu_1 \geq \mu_2 \geq \cdots \geq \mu_n >0$ 
\begin{align}
|Q,\mu \rangle  = \prod_{j} a_{-\mu_j} |Q\rangle, \qquad a_0 |Q,\mu \rangle = Q |Q,\mu \rangle \label{canonical_basis_U(1)}
\end{align}
For multiplicities of element $j$, $ m_j = m_j(\mu)$ the norm is:
\begin{align}
 \langle  Q,\mu |  Q',\mu' \rangle =  z_{\mu}  \delta_{Q,Q'}   \delta_{\mu \mu' },\;\;\;\; z_\mu= \prod _j j^{m_j} m_j! \label{U(1)_scalar_product}
\end{align}
The matrix elements of the primary field between the \emph{normalized} basis of descendants can be computed easily using the recurrence $[ a_m, :e^{i\beta \varphi(z)}:] =    :e^{i\beta \varphi(z)}:  \beta z^m $. They are of the form
 \begin{align}
 \langle  Q', \mu' | :e^{i\beta\varphi(z)}: |Q,\mu \rangle =z^{|\mu'|-|\mu| + \beta Q}A_{\mu',\mu} \delta_{Q',Q+\beta},
\end{align}  
which is consistent with Eq.(\ref{conformalinvarianceandV}) and with charge conservation; $\beta Q$ comes from the difference in conformal dimensions $(Q'^2- Q^2)/2$. With $m_j=m_j(\mu)$ and $m_j' = m_j(\mu')$ being the multiplicities of $j$ in $\mu$ and  $\mu'$, one finds:
 \begin{align}
 A_{\mu',\mu} = \prod_{j\geq 1} \sum_{r =0}^{m_j'}   \sum_{s = 0}^{m_j} \frac{\left(-1\right)^s}{r! s!}  \left(\frac{\beta}{\sqrt{j}}\right)^{r+s}  \delta_{m_j'+s,m_j+r} \frac{\sqrt{m_j'! m_j !}}{(m_j-s)!} \label{Aexplicit}
\end{align} These matrix elements are real (useful for numerics) with the symmetry $A_{\mu,\mu'}(\beta) = A_{\mu',\mu}(-\beta)$.

We now move to the neutral part, starting with the case of $(k,r)=(2,r)$ Jacks. The CFT can be factorized as a $U(1)$ free boson times a neutral minimal model $M(3,2+r)$~\cite{DSM}. The underlying symmetry of this minimal model is the Virasoro algebra 
\begin{align}
[L_n,L_m]= (n-m) L_{n+m}+  \frac{c}{12}n(n^2-1) \delta_{n+m,0} \label{Virasoro_algebra}
\end{align}
 with central charge  $c = 1 - 6(g-1)^2/g$,  $g = \frac{2+r}{3}$. The electron operator is $V(z) = \Psi (z) :e^{i\sqrt{2 h_{\Psi}+m} \varphi (z)}$. $\Psi(z) = \Phi_{(1|2)}(z)$ is a primary field in the neutral CFT, with conformal dimension $h_{\Psi}$, kept generic. Like in the $U(1)$ case, the Hilbert space of the neutral CFT is made of the primary fields $\ket{\Delta}$, eigenstates (with eigenvalue $\Delta$) of $L_0$ and  annihilated by lowering operators $L_n, n>0$, and their descendants,  indexed by a partition $\lambda = (\lambda_1,\lambda_2,\cdots,\lambda_n)$:
\beq
\ket{\Delta, \lambda}  = L_{-\lambda_1} L_{-\lambda_2} \cdots L_{-\lambda_n}| \Delta \rangle,  \;\;\;    \label{Virasoro_canonical_basis}
\eneq
also eigenstates of $L_0$ with eigenvalues $\Delta + |\lambda|$, where $|\lambda| = \sum_i \lambda_i$. Using the Virasoro algebra  \eqref{Virasoro_algebra} and $L_n^{\dagger} = L_{-n}$, we can compute any overlap between descendants.  While descendants with different $|\lambda|$ are clearly orthogonal (having different $L_0$ eigenvalues), a major difference from the $U(1)$ case is that the descendants  \eqref{Virasoro_canonical_basis} are generically independent but not orthogonal. Hence, at each level $|\lambda|$, we have to \emph{numerically} build an orthonormal basis. 

Another issue is that for non unitary CFTs (such as the one underlying the Gaffnian) some descendants have a negative norm.  These states have to be included, and the sign of their norm can be handled by an extra diagonal matrix $D$ acting on the right of the MPS matrices : $B^m \rightarrow B^m D$.

A final major difference from the $U(1)$ case is the presence of null  vectors - states of vanishing norm under the scalar product defined by $L_{n}^{\dagger} = L_{-n}$. This is a reflection of the fact that for special values of  $\Delta$ (which include all the interesting cases), some states in \eqref{Virasoro_canonical_basis} are not independent. CFT characters count the number of independent descendants at each level, from which one can deduce the number of null-vectors. 
Our numerical procedure for computing overlaps reproduces this counting. At each level we need to detect and drop all null-vectors before performing the Gram-Schmidt process.

To compute matrix elements $\langle \Delta' ,\lambda' |  \Psi(1) |\Delta, \lambda \rangle$ between descendants, it is convenient to work in the overcomplete (due to null vectors) "basis"  \eqref{Virasoro_canonical_basis} and then transform back  to the orthonormal basis. The level-$0$ matrix element  $\langle \Delta'|  \Psi(1) |\Delta \rangle $   is simply the OPE structure constant $D_{\Delta',h_{\Psi},\Delta}$, known in the closed form for minimal models and, gives an overall pre-factor which can be ignored. Others can be computed using a similar method as for the $\text{U}(1)$ CFT; for all $m \in \mathbb{Z}$,
\begin{eqnarray}
 [L_{m}-L_0,  \Phi^{(h)}(1) ] =  m\; h \;  \Phi^{(h)}(1), \label{FR_Virasoro}
\end{eqnarray} 
where $\Phi^{(h)}$ is any primary field with a generic conformal dimension $h$. 
Any matrix element can in principle be computed exactly using this method but to the best of our knowledge there is no analytical closed formula. 

We are now in the position to write down the MPS matrices for $(k=2,r)$ Jack states.  The state $| \Delta, \lambda  ; Q, \mu \rangle = | \Delta, \lambda  \rangle \otimes | Q, \mu \rangle$ of an \emph{over-complete} "basis" of states in the CFT for $k=2$ Jack states has level ("momentum") $P = |\lambda| + |\mu|$, which serves as the truncation parameter for the MPS. The matrix elements $\langle \Delta', \lambda' ; Q', \mu' | B^m  | \Delta, \lambda ; Q, \mu \rangle $ for $m=0,1$ are given by:
\begin{align}
B^0 : \;\;\;\; \delta_{\mu,\mu'} \delta_{Q',Q-\frac{1}{\sqrt{q}}}  \delta_{\Delta,\Delta'} \langle \Delta', \lambda' | \Delta, \lambda \rangle \label{B0 Virasoro}
\end{align} 
\begin{align} 
B^1: \;\;\;\;    
  \delta_{\Delta'+|\lambda'|+|\mu'|+\sqrt{q}Q,\Delta +|\lambda|+|\mu|+ h_{\Psi} + 1/2} \nonumber \\ 
\times  \langle \Delta', \lambda' | \Psi(1) |\Delta,\lambda \rangle \,  \delta_{Q',Q+\sqrt{q}-1/\sqrt{q}} \,   A_{\mu',\mu}.
    \label{B1 Virasoro}
\end{align} 
$\langle \Delta', \lambda' | \Psi(1) |\Delta,\lambda \rangle$ can be computed using the neutral CFT, then changed to an orthonormal basis.  $A_{\mu',\mu}$ is given in \eqref{Aexplicit} for $\beta = \sqrt{q}$. The values of $\Delta,\Delta'$ are fixed by the fusion rules of the electron operator in the neutral sector.  For the $(k,r)=(2,r)$ Jacks,  these are $\Psi \times \Psi = 1$, $\Psi \times 1 = \Psi$, which gives rise to two neutral sectors $\ket{x}$ $x=0,1$ with conformal dimension $\Delta = x h_{\Psi}$ and 
\begin{align}
\langle x', \lambda' | \Psi(1) | x ,\lambda \rangle = \delta_{x+x',1} \langle x', \lambda' | \Psi(1) | x,\lambda \rangle.
\end{align}
We have implemented the above MPS numerically and verified that it exactly reproduces all $(k,r)=(2,2), (2,3), (2,6)$ Jack states. The $P=0$ MPS recovers the thin torus limit~\cite{thintorus} (root partition) of these Jacks. The topological sector (responsible for the ground-state degeneracy) can be fixed by choosing matrix elements of the product $B^{m_1} \ldots B^{m_{N_\phi}}$ between different primary fields.  One can describe quasi-hole states by inserting quasi-hole matrices in the MPS, as was done for Laughin and MR states in Ref.~\cite{zaletel-PhysRevB.86.245305}. Or alternatively, edge states are obtained by  choosing matrix elements involving descendant states instead of primaires. This means that the MPS formalism establishes a mapping between edge states and the auxiliary space, which in turns controls the entanglement spectrum. In particular the MPS makes transparent the counting of the orbital entanglement spectrum for such states \cite{Chandran-PhysRevB.84.205136}.

We now move to the Read-Rezayi $Z_3$ state, exemplified by the $k=3$ Jack polynomial. For $k>2$ Jack states, the only known approach is to deal with a CFT with an enlarged algebra, the so-called $\mathcal{W}_k$ algebra (see Eqs. (44), (45), and (46) of Ref.~\cite{estienne-JPhysA-2009}), which includes a $W$ current of spin $3$. This generic approach, which applies to all $(k,r)= (3,r)$ Jack  states becomes inefficient for the $Z_3$ RR state due to the appearance of an extremely large number of null vectors (at each level of truncation). The underlying CFT for the $k=3$ RR state is known to be equivalent to the minimal model  $M(5,6)$, the field $\Psi_{1}(z)$ becoming the primary field $\Phi_{(3|1)}(z)$. This alternative approach provides a basis in which matrix elements involve only Virasoro modes $L_n$.  The electron operator is $ V(z)=  :e^{i\sqrt{q}\varphi (z)}: \, \otimes\, \Psi_{1}(z)$ with  $q = 2/3+ m$. The neutral CFT field  $\Phi_{(3|1)}$ can be split into two chiral fields $\Psi_1(z)$, $\Psi_{-1}(z)$  with conformal dimension $h_{\psi} = 2/3$. Their fusion rules in the $\mathcal{W}_3$ framework are $\Psi_1 \times \Psi_1 = \Psi_{-1} , \Psi_{1} \times \Psi_{-1} = 1 $.  While in the  $\mathcal{W}_3$ algebra language $\ket{W} = \sqrt{3/c} W_{-3} \ket{0}$ is a descendant of the identity, it is primary (with conformal dimension $3$) with respect to the Virasoro algebra: $L_n \ket{W}=0, \;n>0$. Accordingly, in the minimal model $M(5,6)$ framework one has to work with  the fusion rule  $\Psi_1 \times \Psi_{-1} = 1 + W$. 

The $\mathbb{Z}_3$ parafermions have three sectors corresponding to the $\mathbb{Z}_3$ charge of the field $x=0, \pm 1$.  Working in the Virasoro algebra, the $x=0$ sector contains two primaries $\ket{0}, \ket{W}$ as well as their descendants obtained just like above by the action of the Virasoro generators $L_{-n}$, whereas $x =\pm 1$ are made of $\ket{\psi_{\pm 1}}$ and their descendants.   The matrix element between descendants $\langle \Delta',\lambda' | \Psi_1(1) | \Delta,\lambda \rangle$ vanishes unless $x'=x+1$ mod $3$. The matrix elements we need are $\langle \Psi_1,\lambda' | \Psi_1(1) | 0,\lambda \rangle$,  $ \langle \Psi_1,\lambda' | \Psi_1(1) | W,\lambda \rangle$, $ \langle \Psi_{-1},\lambda' | \Psi_1(1) | \Psi_1,\lambda \rangle$, all other being obtained from the above by charge conjugation $\langle \alpha' | \Psi_1(1) | \alpha \rangle = \langle \mathcal{C}(\alpha) | \Psi_{1}(1) | \mathcal{C}(\alpha') \rangle$, where charge conjugation interchanges $\ket{\Psi_{\pm 1}}$, leaves $\ket{0}$ invariant and flips the sign of $\ket{W}$. Using Eq.\eqref{FR_Virasoro}, we can compute these matrix elements up to one coefficient, namely $\bra{\Psi_1} \Psi_1(1) \ket{W}/\bra{\Psi_1} \Psi_1(1) \ket{0}$. But this is simply an OPE structure constant, which is found to be $\sqrt{26}/9$. 

Once the matrix elements (real, with this normalization) between descendants are known, it is easy to find the explicit form of the MPS $B$ matrices for the RR state, $\langle \Delta', \lambda' ; Q', \mu' | B^0  | \Delta, \lambda ; Q, \mu \rangle$ and $  \langle \Delta', \lambda' ; Q', \mu' | B^1  | \Delta, \lambda ; Q, \mu \rangle  $:
\begin{align}
B^0:  \;\;\;\; \delta_{\mu,\mu'} \delta_{Q',Q-1/\sqrt{q}}  \langle \Delta', \lambda' | \Delta, \lambda \rangle \label{B0 W}
\end{align} 
\begin{eqnarray} 
  &  B^1:\;\;\;\;  \delta_{\Delta'+|\lambda'|+|\mu'|+\sqrt{q}Q,\Delta +|\lambda|+|\mu|+ h_{\Psi} + 1/2} \nonumber \\ &
\times  \langle \Delta', \lambda' | \Psi_1(1) |\Delta,\lambda \rangle \,  \delta_{Q',Q+\sqrt{q}-1/\sqrt{q}} \,   A_{\mu',\mu}, \label{B1 W}
\end{eqnarray}
with $h_{\Psi} = 2/3$.   $A_{\mu',\mu}$ is given in \eqref{Aexplicit} for $\beta = \sqrt{q}$.  $\lambda = \{ \lambda_i \}$ is a partition of descendants of all the four primary fields ($W$ included) in the theory. As before, $P= |\lambda| + |\mu|$ is the truncation parameter for the MPS. As before the $P=0$  MPS recovers the root partition $\ldots 1 0^{m-1} 1 0 ^{m-1} 1 0^{m+1} 1 0^{m-1} 1 0^{m-1}1$, which is the thin torus limit of the RR state multiplied by $m$ Jastrow factors.

The obtained MPS description of the Jacks (un-normalized wavefunctions on the annulus) is trivially transmuted to other geometries: for the sphere and the infinite plane, we obtain a site dependent MPS with $B^{m_i}[i] =B^{m_i}/N(i)$, where $N_i$ is the norm of the single particle orbital $z^i$ in the respective geometries. For the cylinder, a site \emph{independent} MPS is possible by introducing a time-evolution $e^{-2 \pi L_0/L}$, with $L$ denoting the circumference of the cylinder. The well-known squeezing properties of FQH states easily follow  from the MPS description. 

\begin{figure}[ttt]
\begin{center}
\includegraphics[width=0.42\textwidth]{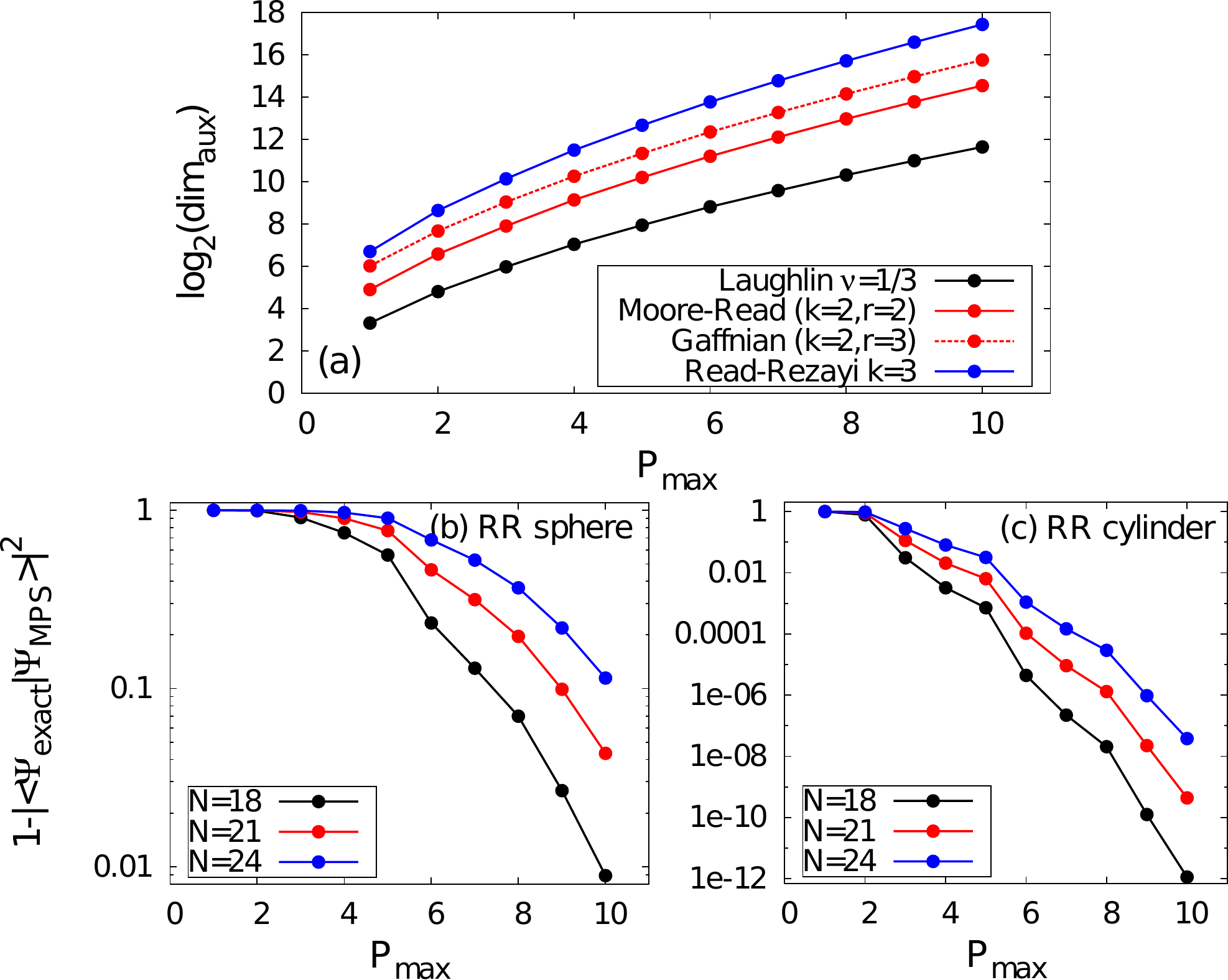} 
 \caption{(Color online). Growth of the auxiliary Hilbert space and convergence criteria for different types of boundary conditions. {\it Upper panel}: Dimension of the auxiliary space for the Laughlin, Moore-Read, Gaffnian and Read-Rezayi $Z_3$ state as a function of $P_{\rm{max}}$. {\it Lower panel} : Quantum distance between the exact Read-Rezayi $Z_3$ and the truncated MPS state on the sphere (b) and on the cylinder with aspect ratio one (c) as a function of $P_{\rm{max}}$. Note that for $N=24$ and $P_{\rm{max}}=10$, the MPS only gives 40\% of the $6.2\times 10^7$ components. Smaller quantum distances are obtained for Laughlin, Moore-Read and Gaffnian (not shown). \label{mpsfigure} } 
\end{center}
\end{figure}

So far the MPS description we have provided is exact. For numerical purposes of approximating a given state, we introduce the truncation level $P_{\rm{max}}$, which is the maximum allowed value of $P=|\lambda| + |\mu|$. $P_{\rm{max}}= 0$ gives the root partition (thin-torus limit) of the Jack polynomials and amounts to dropping all descendants in the CFT Hilbert space. $P_{\rm{max}}= 1$ gives the correct weights for all configurations obtained through a \emph{single} squeezing of the root partition. Generically, the truncation at $P_{\rm{max}}$ amounts to restricting the number of any squeezing steps from the root partition to $P_{\rm{max}}$. This is also the momentum quantum number labeling the entanglement spectrum levels~\cite{li2008}. Due to the shape of the orbital spectrum, the truncation to a certain $P_{\text{max}}$ is expected to be equivalent to keeping the states with the highest Schmidt weight in the groundstate. This is also related (though not equivalent) to expansions around the thin cylinder limit~\cite{nakamura,soule-PhysRevB.85.155116}, where the weights of configurations decrease with the amount of squeezings from the root partition (equivalent to the exponential decay of correlation functions in the associated CFT).

In Fig.~\ref{mpsfigure} we provide numerical benchmarks for the accuracy of approximating the full Jack states by the MPS truncated at level $P_{\text{max}}$. The approximate Laughlin, Moore-Read, Gaffnian and Read-Rezayi states have been constructed by MPS $B$ matrices whose auxiliary Hilbert space dimension grows as shown in Fig.~\ref{mpsfigure}(a). The accuracy of the approximation is quantified by the overlap of an MPS state with a full Jack polynomial, and an example for the Read-Rezayi $Z_3$ state is given in Figs.~\ref{mpsfigure}(b,c). We observe that  the approximate MPS state becomes an excellent approximation of the exact FQH state for relatively low values of $P_{\text{max}}=10$. Note that the convergence to the exact state on the sphere [Fig.~\ref{mpsfigure}(b)] is strikingly slower than on the cylinder [Fig.~\ref{mpsfigure}(c)], making this a preferred type of boundary condition for DMRG implementations~\cite{Shibata-PhysRevLett.86.5755,Feiguin-PhysRevLett.100.166803,donna,dmrghu}. 

In conclusion, we have provided a method to obtain the MPS description of FQH model states given by correlators of the primary fields in a CFT. We have furthermore obtained the exact MPS form of the $(k,r)=(2,r), (3,2)$ Jack states (including Moore-Read, Gaffnian, and Read-Rezayi $k=3$ state), and compared the approximate MPS (truncated to a certain $P_{\text{max}}$) with the exact states. Comparatively small values of $P_{\text{max}}$ were found to be sufficient for obtaining extremely accurate approximations of these states on the cylinder, which might have consequences for the improved DMRG implementations of realistic (Coulomb) Hamiltonians.

We wish to thank F.D.M.~Haldane, A. Sterdyniak, R.Santachiara and J.~Dubail for inspiring 
discussions. BAB and NR were supported by NSF CAREER DMR-095242, ONR-N00014-11-1-0635, 
ARMY-245-6778, MURI-130-6082, Packard Foundation, and Keck grant. ZP acknowledges support
by DOE grant DESC0002140 and thanks KITP for hospitality (supported in part by NSF PHY11-25915).

 \bibliography{mpsshort}

\end{document}